\documentclass[12pt]{iopart}
\usepackage{hyperref}
\usepackage{graphicx}
\usepackage{iopams}
\usepackage{bm}
\usepackage{cite}

\def\harddimer{hard-dimer}
\def\harddimers{hard dimers}

\begin{document}
\title{Magnetocaloric effect in one-dimensional antiferromagnets}

\author{M. E. Zhitomirsky}
\address{Commissariat \`a l'Energie Atomique, DSM/DRFMC/SPSMS,
17 avenue des Martyrs, 38054 Grenoble, C\'edex 9 France}

\author{A. Honecker}
\address{Technische Universit\"at Braunschweig, Institut f\"ur 
Theoretische Physik, Mendelsohnstrasse 3, 38106 Braunschweig, Germany}

\vspace{16pt plus3pt minus3pt}
\begin{indented}
\item[]\rm 30 April 2004; revised 8 July 2004 
\end{indented}

\begin{abstract}
An external magnetic field induces large relative changes in the entropy 
of one-dimensional quantum spin systems at finite temperatures. 
This leads to a magnetocaloric effect, {\em i.e.}\ a change in 
temperature during an adiabatic (de)mag\-ne\-tization process. Several examples 
of one-dimensional spin-1/2 models are studied by employing the Jordan-Wigner 
transformation and exact diagonalization. 
During an adiabatic (de)magnetization
process temperature drops in the vicinity of a field-induced
zero-temperature quantum phase transition. Comparing different levels of
frustration, we find that more frustrated systems cool down to lower
temperatures. For geometrically frustrated spin models a finite
entropy survives down to zero temperature at certain magnetic fields.
This property suggests frustrated quantum spin systems
as promising alternative refrigerant materials for low-temperature
magnetic refrigeration.
\end{abstract}


\pacs{75.30.Sg,	
      75.10.Jm,	
      75.50.Ee	
}

\section{Introduction}

The magnetocaloric effect, or heating/cooling of a magnetic system
under an adiabatic (de)magnetization process, is related to the isothermal
variation of the entropy in a magnetic field via a simple thermodynamic
relation
\begin{equation}
\left(\frac{\partial T}{\partial H}\right)_S =
- T\,\frac{(\partial S/\partial H)_T}{C} \ ,
\label{dTdH}
\end{equation}
where $C$ is the specific heat at a constant magnetic field.
A field variation of entropy can be found via a temperature
variation of the magnetization:
$(\partial S/\partial H)_T = (\partial M/\partial T)_H$.
Standard examples of magnetic materials with a significant magnetocaloric 
effect include paramagnetic salts, which consist of noninteracting spins, 
and ferromagnets near the Curie point, where $(\partial M/\partial T)_H$ 
is, generally, large \cite{cyrot,tishin}.
Besides a fundamental interest, the magnetocaloric effect has 
a great importance for the technique of magnetic cooling, 
which has been employed over the years to reach 
temperatures in a sub-Kelvin range \cite{lounasmaa}.
A certain progress has also been achieved to utilize this 
technique for room temperature refrigeration \cite{pecharsky,tegus}.

Recently, it has been predicted that an enhanced magnetocaloric effect 
exists in the vicinity of finite-field transitions in a class of geometrically 
frustrated antiferromagnets \cite{mzh03}. Local spins in such magnets 
reside on special types of crystal lattices, with pyrochlore, kagome, and 
garnet lattices being a few well-known examples. Competing antiferromagnetic 
interactions on frustrated lattices lead to a {\it macroscopic} degeneracy
(entropy) of classical ground states. Large variations of the total entropy 
occur near the saturation field $H_c$, where the transition to a fully 
polarized nondegenerate state takes place. Above $H_c$ in the saturated phase, 
highly frustrated magnets have a flat dispersionless low-energy branch of 
magnons with energy $(H-H_c)$. Dispersionless excitations can be represented
in real space as localized quasiparticles. Such localized magnon excitations
have recently also been studied in frustrated quantum spin models
\cite{schulenburg,richter} where it was shown that
the corresponding many-particle states are not only exact eigenstates of
the Hamiltonian, but indeed the lowest ones. Hence, the
macroscopic zero-$T$ entropy of geometrically frustrated magnets at
$H=H_c$ is robust against quantum fluctuations \cite{RSH} (see also \cite{DeRi}), 
whereas below $H_c$ the quantum (or thermal) order by disorder effect should lift 
the classical degeneracy \cite{schulenburg,ZhHP}. 
It is the condensation of a  macroscopic number of localized magnons 
near the saturation field, which leads to the enhanced magnetocaloric
effect in geometrically frustrated magnets \cite{mzh03}.

The aim of the present work is two-fold. First, we argue that a large 
magnetocaloric effect appears near continuous phase transitions in a magnetic
field for general one-dimensional (1D) quantum spin systems. 
This phenomenon has a simple physical origin. The excitation spectrum
at the quantum critical point is different and, generally, softer
if compared to spectra above and below the transition.
The density of low-energy modes is strongly enhanced 
in spin chains due to the 1D van Hove singularity. 
Therefore, the soft modes increase the total magnetic entropy 
around the transition point and produce a sizeable magnetocaloric effect.
We present analytic calculations for two exactly solved models: 
an $XY$ antiferromagnetic chain with a magnetic field applied perpendicular
to the plane and an Ising model in a transverse field.
Then, we study by means of exact diagonalization of finite lattices
a spin-1/2 Heisenberg chain.
Secondly, we investigate how the magnetocaloric effect in 1D
systems changes with increasing degree of frustration.
For this purpose we consider a frustrated $J_1$--$J_2$ spin-1/2 chain
and the so-called sawtooth spin chain
as a 1D toy model for the class of geometrically frustrated lattices
containing the kagome and pyrochlore lattice.

\section{$XY$ chain in transverse field}

We begin with a simple spin-1/2 $XY$ antiferromagnet in a transverse magnetic 
field, which is described by the Hamiltonian:
\begin{equation}
\hat{\cal H}= \sum_{i=1}^N \left[ J\left(S_i^x S_{i+1}^x
+  S_i^y S_{i+1}^y\right) - H S_i^z \right] .
\label{HXY}
\end{equation}
With the help of the Jordan-Wigner transformation \cite{takahashi} this
is mapped to a Hamiltonian of free spinless fermions:
\begin{equation}
\hat{\cal H}=  - \frac{H\,N}{2} + \sum_k \varepsilon_k 
c^\dagger_k c_k \ , \ \ \ \ \varepsilon_k = H+J\cos k \ .
\label{Hdiag}
\end{equation}
In high magnetic fields the ground state
is the fully polarized spin state, which describes  the fermion vacuum.
As the magnetic field is lowered, the excitation gap for fermions
decreases and vanishes at the saturation field
$H_c = J$. At the transition point 
the low-energy fermions have a parabolic dispersion
$\varepsilon_{\pi+q} \approx \frac{1}{2} Jq^2$ for $|q|\ll 1$.

The  heat capacity of a gas of free fermions is
\begin{equation}
C = \frac{1}{T^2} \sum_k \varepsilon_k^2 n_k(1-n_k) \ ,
\end{equation}
where $n_k = [{\rm e}^{\varepsilon_k/T} + 1]^{-1}$ are the occupation numbers.
At the saturation field the specific heat has a square-root low-temperature
dependence
\begin{equation}
C/N = \frac{3}{4}\:\zeta\left({\textstyle\frac{3}{2}}\right)
\left(1-\frac{1}{\sqrt{2}}\right) \sqrt{\frac{T}{2\pi J}} 
\label{CHc}
\end{equation}
with $\zeta(\frac{3}{2})\approx 2.612375$.

The magnetization of an $XY$ chain is given by
$M = \frac{1}{2}N - \sum_k n_k$,
which reduces at zero temperature to 
\begin{equation}
M/N = \frac{1}{2} - \frac{1}{\pi} \arccos(H/H_c)  
\end{equation}
(compare with eq.~(3.5) of \cite{Katsura}).
The magnetization curve $M(H)$ has a standard square-root singularity 
near the saturation field \cite{DzNe,PoTa}.
The magnetocaloric effect is calculated as 
\begin{equation}
\left(\frac{\partial S}{\partial H}\right)_T =  
 - \frac{1}{T^2} \sum_k \varepsilon_k n_k(1-n_k) \ .
\end{equation}
At $H=H_c$ and low temperatures one finds
\begin{eqnarray}
&& \left(\frac{\partial S}{\partial H}\right)_T = 
-\frac{c_1N}{2\sqrt{2\pi TJ}} \ , \nonumber \\
&& c_1=\frac{2}{\sqrt{\pi}}\:\int_0^\infty \frac{dx}{{\rm e}^{x^2}+1} =
\sum_{n=1}^{\infty}\frac{(-1)^{n-1}}{n^{1/2}}\approx 0.604899 \  .
\label{dSHc}
\end{eqnarray}

The low-temperature expressions for the entropy above, at and below
the saturation field $H_c$ are given by
\begin{eqnarray}
S/N = \frac{\Delta_H}{\sqrt{2\pi T}}\: {\rm e}^{-\Delta_H/T}\ ,&  &H>H_c \ , 
\nonumber \\
S/N = \frac{3}{2}\:\zeta\left({\textstyle\frac{3}{2}}\right) 
\left(1-2^{-1/2}\right) \sqrt{\frac{T}{2\pi J}}\ , &\qquad\quad  
& H=H_c \ , \label{eqSxy} \\
S/N = \frac{\pi}{3} \frac{T}{\sqrt{2|\Delta_H|J}}\ ,  && H<H_c \ ,
\nonumber
\end{eqnarray}
where $\Delta_H = H-H_c$ and $|\Delta_H|\gg T$ is assumed.
At $T=0$ in an arbitrary magnetic field the system has a unique quantum ground 
state and, correspondingly, zero total entropy. At $T>0$, thermal 
fluctuations yield different entropy contributions in the above three
regimes. Variation of the applied magnetic field leads, therefore, to large
relative changes in the total entropy and to a large
magnetocaloric effect.

\begin{figure}[t]
\begin{center}
\includegraphics[width=0.7\columnwidth]{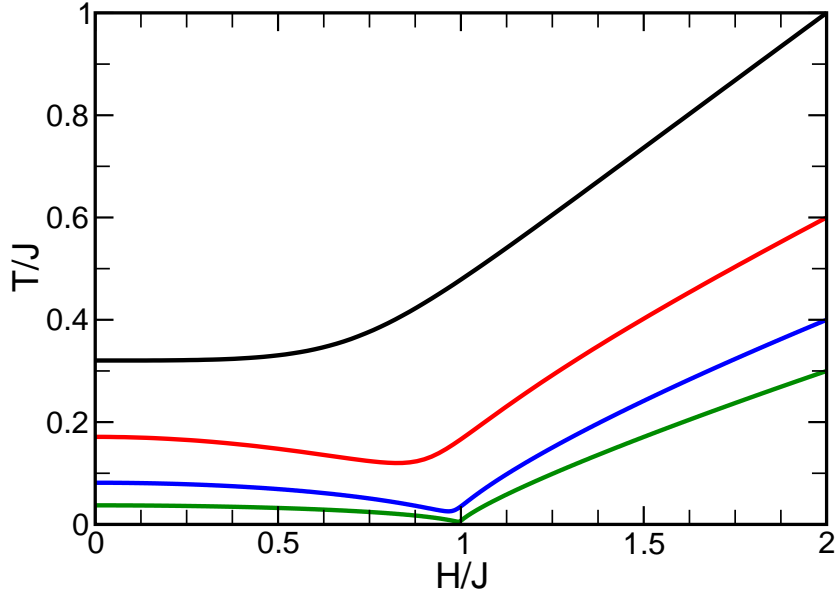}
\end{center}
\caption{\label{XYad} 
Adiabatic demagnetization curves of the $XY$ antiferromagnetic $S=1/2$ chain.}
\end{figure}

The adiabatic demagnetization curves of an $XY$ antiferromagnet
can be found either from a direct numerical solution of 
$S(H,T) ={\rm const}$   
or by numerical integration of the differential relation (\ref{dTdH}).
The results are presented in Fig.~\ref{XYad} for $H_i=2H_c$ and several
starting temperatures $T_i$. 
At $H=H_c$ the cooling rate
$(\partial T/\partial H)_S$ has the maximal value determined by
eqs.~(\ref{dTdH}), (\ref{CHc}) and (\ref{dSHc}).
The lowest temperature of
an adiabatic process is reached at $H^*<H_c$.
At low temperatures the difference between these two fields
becomes, however, very small.
Using eqs.~(\ref{eqSxy}) we find the following analytic expressions
for a field-dependence of temperature. At $(H-H_c)\gg T$, temperature changes
nearly linearly with 
$(H-H_c)$ due to a dominant exponential contribution. In this region
\begin{equation}
\frac{1}{T} \left(\frac{\partial T}{\partial H}\right)_S=\frac{1}{H-H_c} \, .
\label{aboveHc}
\end{equation}
The lowest temperature reachable during an adiabatic demagnetization process 
from $T_i$ and $H_i>H_c$ to $H_f=H_c$ is exponentially small:
\begin{equation}
T_f \sim \frac{(H_i-H_c)^2}{T_i}
\exp\left[-\frac{2(H_i-H_c)}{T_i}\right] .
\label{eqTf}
\end{equation}
Below the saturation field, temperature decreases as $(H_c-H)^{1/2}$ 
with $H\to H_c$ [$(H_c-H)\gg T$]. 
In this region one finds
\begin{equation}
\frac{1}{T} \left(\frac{\partial T}{\partial H}\right)_S =
   \frac{1}{2} \, \frac{1}{H-H_c} \, .
\label{belowHc}
\end{equation}
The lowest temperature reachable
during adiabatic magnetization from $H=H_i<H_c$ to $H_f=H_c$ is
\begin{equation}
T_f \sim \frac{T_i^2}{H_c-H_i} \ .
\end{equation}

\section{Heisenberg chain in magnetic field}

We now discuss the effect of $S^z$-$S^z$ interaction on the
magnetothermodynamics of the phase transition in the spin-1/2 $XXZ$ chain:
\begin{equation}
\hat{\cal H}= \sum_i \left[ J \left(S_i^x S_{i+1}^x + S_i^y S_{i+1}^y
+ \Delta S_i^z S_{i+1}^z\right) - H S_i^z \right] .
\label{HHeis}
\end{equation}
The case $\Delta=1$ corresponds to the Heisenberg chain.
The interaction of the $z$-components shifts the fermion energies
and the saturation field:
\begin{equation*}
\varepsilon_k = H - J\Delta + J\cos k\ , \ \ \ \ \
H_c = J(1 + \Delta) \ .
\end{equation*}
It also induces an interaction between fermions:
\begin{eqnarray}
V &=& J\Delta \sum_i c_i^\dagger c_i c_{i+1}^\dagger c_{i+1}  \nonumber  \\
  &=& \frac{1}{2N} \sum_{k,k',q} J\Delta[\cos q - \cos(k-k'+q)] 
c_{k+q}^\dagger c_{k'-q}^\dagger c_{k'} c_k \ .
\end{eqnarray}
The interaction between spinless fermions vanishes as $k,k',q\rightarrow 0$.
In the vicinity of the saturation field only wave-vectors near $k\approx \pi$
are important and, therefore, interactions play a minor role
\cite{DzNe,PoTa,sachdev}.
The main correction to the free energy at low temperatures is the Hartree-Fock
contribution, which is estimated as
\begin{equation}
\delta F_{\rm HF} = \langle V \rangle = JN\Delta (n^2-m^2) 
\end{equation}
with $n = \langle c_i^\dagger c_i\rangle$ and 
$m = \langle c_i^\dagger c_{i+1}\rangle$. 
At the saturation field one finds
\begin{eqnarray*}
n - m & = & \sum_k \frac{1-\cos k}{{\rm e}^{\varepsilon_k/T}+1} = 
   2 c_1\, \sqrt{\frac{T}{2\pi J}} \ , \\
n + m & = & \sum_k \frac{1+\cos k}{{\rm e}^{\varepsilon_k/T}+1} = 
\frac{1}{2}\,\sqrt{\frac{T^3}{2\pi J^3}}\:%
\zeta\left({\textstyle\frac{3}{2}}\right)
\left(1-2^{-1/2}\right) . 
\end{eqnarray*}
The temperature expansion of the free energy at $H=H_c$ is
\begin{equation}
F/N  =  - \zeta\left({\textstyle\frac{3}{2}}\right)\left(1-2^{-1/2}\right)
\left[\frac{T^{3/2}}{(2\pi J)^{1/2}}-\Delta c_1\:
\frac{T^2}{2\pi J}\,\right],
\label{Fheis}
\end{equation}
where the first $T^{3/2}$ term is a Fermi-gas contribution 
and the second term $T^2$ is the Hartree-Fock correction.
The next order $T^{5/2}$ correction to eq.~(\ref{Fheis}) 
is determined by deviations
from a parabolic dispersion in $\varepsilon_k$.
The interaction gives a negligible contribution in the thermodynamics 
of a Heisenberg chain at temperatures $T\lesssim 0.1J$.
The expression (\ref{Fheis}) agrees with the exact thermodynamic Bethe ansatz
solution for a Heisenberg chain \cite{takahashi,takahashi73},
which, however, gave the second $T^2$ term without numerical
prefactor.

\section{Frustrated Heisenberg chain near saturation}

The density of low-energy states at the saturation field
can be increased by frustration. As a first example we consider
the $J_1$--$J_2$ chain with exchange interaction between
the first and second neighbours at the point $J_2 = \frac{1}{4}J_1$,
where the dispersion degenerates from the generic quadratic
behaviour to a fourth power, giving rise to a fourth-root instead
of the usual square-root singularity in the $T=0$ magnetization
curve at the saturation field \cite{SGMK,GMK,zigzag}. 

The $J_1$--$J_2$ Heisenberg chain is described by the Hamiltonian
\begin{equation}
\hat{\cal H}= \sum_i \left[ J_1 {\bf S}_i\cdot {\bf S}_{i+1} + 
J_2 {\bf S}_i\cdot {\bf S}_{i+2} - H S_i^z \right] \, . 
\label{HFrust}
\end{equation}
The fermion energies in the saturated phase are given by
\begin{equation}
\varepsilon_k = H + J_1(\cos k-1) + J_2(\cos 2k-1) \ .
\label{frus1P}
\end{equation}
Frustrated interaction with the second neighbours does not change 
the saturation field  value for $J_2\leq  \frac{1}{4}J_1$:  $H_c=2J_1$. 
The long-wave length expansion of $\varepsilon_k$ near the minimum point yields
\begin{equation*}
\varepsilon_{\pi+q}\approx H-H_c + \left(\frac{J_1}{2}-2J_2\right) q^2 
+ \left(\frac{2J_2}{3} - \frac{J_1}{24}\right) q^4 \, .
\end{equation*}
At the point $J_2=\frac{1}{4}J_1$, the first $q^2$-term vanishes
and the dispersion of fermions is described by a quartic parabola: 
$\varepsilon_{\pi+q} \approx H-H_c + \frac{1}{8}Jq^4$ ($J_1\equiv J$).
This leads to a higher density of states above the gap
$\rho(\omega) \sim \omega^{-3/4}$ than for the standard 1D van Hove 
singularity $\rho(\omega) \sim \omega^{-1/2}$. For a free fermion gas
with a general dispersion law $\varepsilon_k = A k^z$, the entropy
behaves as
\begin{equation}
S/N = \frac{1+z}{\pi z}\: 
\Gamma\left(1+{\textstyle\frac{1}{z}}\right)
\zeta\left(1+{\textstyle\frac{1}{z}}\right)
\left(1-2^{-1/z}\right)\frac{T^{1/z}}{A^{1/z}} \ .
\label{eqLTalpha}
\end{equation}
For $z=4$ this gives $S\sim T^{1/4}$. However,
for $z>3$ the interaction between fermions becomes again
a relevant perturbation at $H=H_c$ \cite{sachdev} such that 
eq.~(\ref{eqLTalpha}) is no longer expected to give the exact
low-temperature asymptotics. 
Thermal fluctuations tend to reduce frustration by generating
a temperature-dependent quadratic
term $\varepsilon_k \approx B(T)k^2 + A k^z$, which has to be 
determined self-consistently.
In any case, we expect
that at low temperatures the entropy of the frustrated chain is
higher than the entropy of a Heisenberg chain.  Since above $H_c$
the entropy has a similar exponential behaviour given by eq.~(\ref{eqSxy}),
the frustrated chain has larger relative variations
of the entropy and, consequently, should exhibit a larger magnetocaloric 
effect.

\section{Numerical results at the saturation field}

In this section we compare the analytic results for the $S=1/2$ Heisenberg
chain to numerical results obtained by exact diagonalization of
finite rings and check  the scaling of the entropy
for the $J_1$--$J_2$ Heisenberg model.
A spectral representation is used for the evaluation of the thermodynamic
quantities. In particular, the entropy is computed using
\begin{equation*}
S = {1 \over T\, Z} \sum_n E_n \, {\rm e}^{-E_n/T} + \ln Z \, ,
\end{equation*}
where $Z = \sum_n {\rm e}^{-E_n/T}$ is the partition function.
The eigenvalues $E_n$ are obtained by full diagonalization of the Hamiltonian.

Since the CPU time for a full diagonalization grows as the third power of
the dimension, it is
important to use symmetries to reduce the dimension. We have used
conservation of $S^z$, the $z$-component of the total spin, spin-inversion
symmetry for $S^z=0$ and $SU(2)$-symmetry to reconstruct the spectra
in the sector $S^z=1$. Then the biggest dimension arises for $S^z=2$.
The actual size depends on the spatial symmetries, namely translations
and reflections, which have also been exploited. Full
diagonalizations have been performed with up to $N=20$ sites for all
three spin-1/2 systems to be discussed below. The sawtooth chain
has lowest spatial symmetry, hence the largest dimension occurs here, namely
12 618 for $S^z = 2$ and $N=20$. 
Since the magnetic field $H$ couples to the total magnetization $S^z$, which is a conserved
quantity, one computation yields the dependence of the entropy $S$ both
on temperature $T$ and the magnetic field $H$.

\begin{figure}[t]
\begin{center}
\includegraphics[width=0.7\columnwidth]{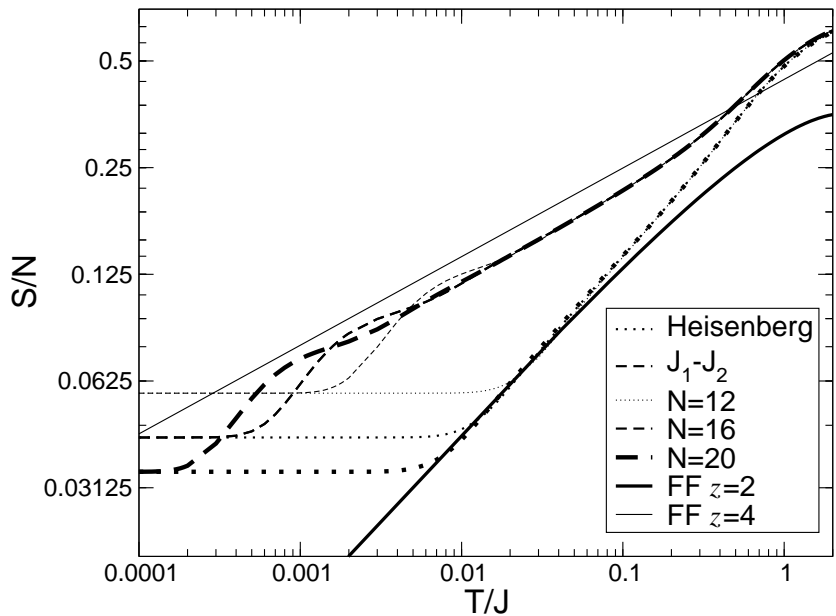}
\end{center}
\caption{\label{numHsat}
Numerical results for the
entropy of the spin-1/2 Heisenberg chain (dotted lines) and the $J_1$--$J_2$ chain
with $J_1=J$, $J_2=\frac{1}{4}J$ (dashed lines) at the saturation field $H_c = 2J$.
The behaviour for a free fermion gas (FF) is shown by the bold solid line, 
eq.\ (\ref{Eheis}), and the thin solid line, 
eq.\ (\ref{eqLTalpha}) with $z = 4$ and $A=\frac{1}{8}J$.}
\end{figure}

Fig.~\ref{numHsat} presents a log-log plot of the temperature-dependence
of the entropy of the Heisenberg chain and the $J_1$--$J_2$ chain with
$J_1=J$, $J_2=\frac{1}{4}J$ at the saturation field $H_c=2 J$ for $N=12$, $16$,
and $20$. At low temperatures, the ground state at $H_c$ is two-fold
degenerate, {\it i.e.}\ $S=\ln 2$ for any $N$ as $T \to 0$. Accordingly,
one observes finite-size effects in $S/N$ at low temperatures. At high 
temperatures, all curves recover the full entropy $S/N = \ln 2 \approx 0.6931$
of a spin-1/2 system. Finite-size effects are still negligible
at intermediate temperatures and the numerical curves
are straight lines in the log-log representation of Fig.~\ref{numHsat},
implying a power law for $S/N$ as a function of temperature. Next, we
discuss these power laws in more detail and argue that the exponents
are indeed the expected ones.

The bold solid line in Fig.~\ref{numHsat} obtained from eq.~(\ref{Fheis}) is
(see also eqs.~(\ref{eqSxy}) and (\ref{eqLTalpha}))
\begin{equation}
S/N = 0.457874\, (T/J)^{1/2} - 0.1473254\,(T/J) \ .
\label{Eheis}
\end{equation}
This fits the numerical data nicely in a wide temperature range 
$0.007 \lesssim T/J \lesssim 0.2$. Deviations from (\ref{Eheis})
for lower temperatures are determined by finite-size effects
in the numerical data, whereas at higher temperatures
further corrections to (\ref{Eheis}) become important. Although indeed
the agreement at higher temperatures can be improved by including
the next correction $O(T^{3/2})$, we will not pursue this further here.

The thin sold line in Fig.~\ref{numHsat} shows eq.~(\ref{eqLTalpha}) with
$z=4$ and $A=\frac{1}{8}J$ appropriate for the frustrated chain. Comparison
with the numerical data for intermediate temperatures confirms
the expected power-law behaviour $S\sim T^{1/4}$. However, the prefactor
derived from eq.~(\ref{eqLTalpha}) for $z = 4$ is by about 20\% too big
in contrast to the Heisenberg chain ($z=2$). This behaviour
of the prefactor reflects the fact that interactions are relevant
for $z = 4$ and partially reduce the entropy compared to
the free-gas expression eq.~(\ref{eqLTalpha}).

\section{Anisotropic $XY$ chain in transverse field}

Low crystal symmetry of real (quasi) one-dimensional magnetic materials
is compatible with various types of spin anisotropy.
If the magnetic field is applied in a general direction, the anisotropy
leads to a nontrivial behaviour by
breaking the conservation of the total magnetization.
It is, therefore, important to investigate the effect 
of spin anisotropy on the magnetocaloric effect.
The simplest kind of anisotropy is anisotropic exchange interaction.
We consider the following Hamiltonian
\begin{equation}
\hat{\cal H}= \sum_i \left( J_xS_i^x S_{i+1}^x  + J_yS_i^y S_{i+1}^y 
- H S_i^z \right) ,
\label{Hanis}
\end{equation}
which at $J_x=J_y$ reduces to the $XY$ chain and at $J_y=0$ describes an Ising
chain in a transverse field.
Introducing new parameters $J=\frac{1}{2}(J_x+J_y)$ and 
$\gamma = (J_x-J_y)/(J_x+J_y)$ and applying the 
Jordan-Wigner transformation, the Hamiltonian (\ref{Hanis})
is reduced to 
\begin{equation}
\hat{\cal H}= -\frac{H\,N}{2}\! +\! \sum_k (H\! +\! J\cos k) c_
k^\dagger c_k
+ \frac{i\gamma J}{2} \sin k (c_k^\dagger c_{-k}^\dagger\! - c_{-k} c_k) \, .
\label{Hisk}
\end{equation}
The Bogoliubov transformation from original fermion
operators $c_k$ to new operators $a_k$:
\begin{equation*}
c_k = u_k a_k + iv_k a_{-k}^\dagger \ , \ \ \ 
c_{-k}= u_k a_{-k} - iv_k a_k^\dagger
\end{equation*}
diagonalizes the quadratic Hamiltonian (\ref{Hisk})
under the two conditions: (i)  $u_k^2+v_k^2=1$ and  (ii)
\begin{equation*}
2u_k v_k = - \frac{\gamma J\sin k}{\sqrt{(H+J\cos k)^2 
+ \gamma^2J^2 \sin^2 k}} \ .
\end{equation*}
The transformed Hamiltonian has a simple diagonal form:
\begin{equation}
\hat{\cal H} = \sum_k 
      \varepsilon_k \left( a_k^\dagger a_k - \frac{1}{2}\right) \, ,
\quad
\varepsilon_k = \sqrt{(H+J\cos k)^2 + \gamma^2 J^2 \sin^2 k} \, .
\end{equation}

\begin{figure}[t]
\begin{center}
\includegraphics[width=0.7\columnwidth]{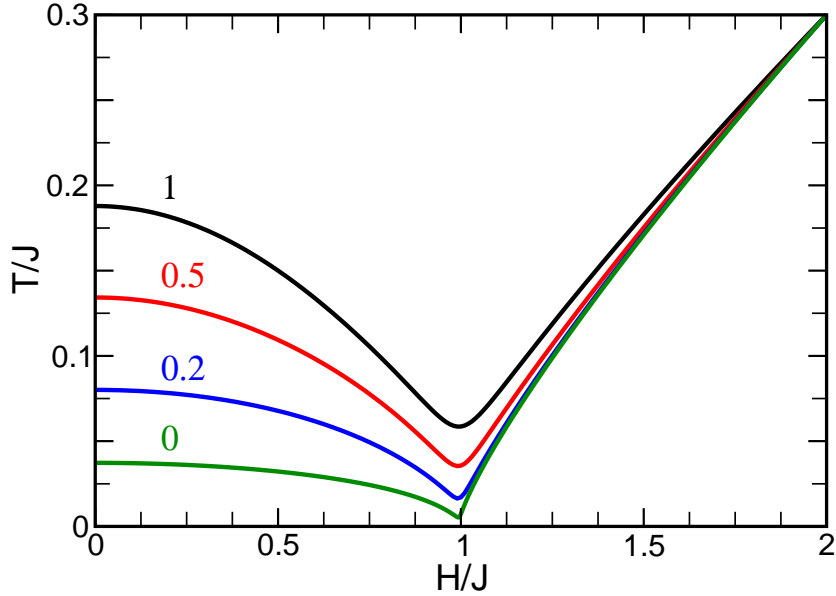}
\end{center}
\caption{\label{AnisXY} 
Dependence of the magnetocaloric effect on anisotropy.
The number near each curve indicates the corresponding value of the
anisotropy parameter $\gamma$.}
\end{figure}

The anisotropic $XY$ chain (\ref{Hanis}) has a phase transition at
$H_c=J$. Near the transition point the long-wavelength
spectrum is given by ($\Delta_H = H-H_c$)
\begin{equation*}
\varepsilon_{\pi+q} \approx \sqrt{\Delta_H^2 + \gamma^2 J^2 q^2} \ .
\end{equation*}
The dispersion acquires a relativistic form at $H=H_c$: 
$\varepsilon_{\pi+q} \approx \gamma J q$. 
The expressions for the free energy, entropy and the specific heat
are the same as for a gas of free fermions. The magnetization 
of the anisotropic $XY$ chain is 
\begin{equation}
M = \sum_k \frac{H+J\cos k}{\varepsilon_k} \left(\frac{1}{2} - n_k\right) .
\end{equation}
At zero temperature $n_k\equiv 0$ and  the total
magnetization does not saturate above $H_c$. 
In the case of the Ising chain $\gamma=1$,
the magnetization has a universal value $M(H_c) = 1/\pi$ at $T=0$. 
The square-root singularity near $H_c$ is replaced with a weaker
logarithmic behaviour: $\partial M/\partial H \sim \ln(\pi\gamma/\Delta_H)$.

The expression for the magnetocaloric effect changes compared
to the $XY$ chain
\begin{equation}
\left(\frac{\partial S}{\partial H}\right)_T = - \frac{1}{T^2}
\sum_k (H+J\cos k) n_k(1-n_k) \, .
\end{equation}
Due to a linear dispersion
law the density of low-energy states in the anisotropic chain remains finite 
instead of the square-root divergence typical for isotropic 
spin chains.
As a result, the entropy at $H=H_c$ depends linearly on temperature:
\begin{equation}
S = \frac{\pi T}{6\gamma J} \ .
\label{Z2Shc}
\end{equation}
This leads to a reduction of the magnetocaloric effect compared to the isotropic
case. The cooling rate at the saturation field also changes its temperature
dependence for the anisotropic chain:
\begin{eqnarray}
&& \left(\frac{\partial T}{\partial H}\right)^{H=H_c}_S = \frac{T}{2\gamma^2}
\ \ \ \ \ \ \ \ \ \ \ \ \ \ \ \ \ \ \ \ \,{\rm anisotropic}\ XY \ , 
\label{dTdHcZ2} \\
&& \left(\frac{\partial T}{\partial H}\right)^{H=H_c}_S =  
\frac{2c_1}{3\zeta(\frac{3}{2})(1-\frac{1}{\sqrt{2}})}
\ \ \ \ \ \ \ {\rm isotropic}\ XY \ .
\label{dTdHcU1}
\end{eqnarray}
The dependence of the adiabatic demagnetization process on the anisotropy
is shown in Fig.~\ref{AnisXY}.
The lowest temperature reachable at $H\approx H_c$ 
grows significantly from the isotropic case
$\gamma=0$ to the strongest anisotropy $\gamma=1$. Nevertheless, even
an Ising chain exhibits sizeable variations of temperature.
Another difference with the isotropic chain is a stronger temperature
increase below $H_c$. This is related to a gap opening in the anisotropic
chain, which leads to a significant reduction of entropy.

\section{Universal properties}

\label{secUni}

Universal low-temperature scaling at second-order quantum phase transitions
is related to the 
dispersion of gapless excitations at the transition point.
In particular, the scaling laws (\ref{dTdHcZ2}), (\ref{dTdHcU1}) remain valid
in the general case:
$(\partial T/\partial H)^{H=H_c}_S = O(1)$ for $U(1)$-symmetry about
the field direction and $(\partial T/\partial H)^{H=H_c}_S = O(T)$ for
$\mathbb{Z}_2$-symmetry about the field. In fact, this is not only true
close to the saturation field, but also in the vicinity of plateaux in
the magnetization curve \cite{RSH,MiKo,habil}. Since these field-induced
transitions are universal (at least in one dimension), the scaling laws 
given by eqs.\ (\ref{dTdHcZ2}) and (\ref{dTdHcU1})
should apply to generic second-order transitions in one dimension.

Our results can be related to a general scaling analysis of
the magnetocaloric effect close to quantum critical points \cite{rosch}.
First we note that e.g.\ the power laws in eqs.\ (\ref{CHc}), (\ref{eqSxy}),
(\ref{eqLTalpha}) and (\ref{Z2Shc}) can be written as
$C$, $S \propto T^{d/z}$ at $H=H_c$ and
$C$, $S \propto T^{y_0} \left|H-H_c\right|^{\nu (d - y_0 z)}$
for $H \ne H_c$, using the notations of \cite{rosch}.
The case of $U(1)$-symmetry is described by the exponents ($d=1$)
\begin{equation}
z=2 \, , \qquad \nu=\frac{1}{2} \, ,
\label{expoU1}
\end{equation}
$y_0 = 1$ in a gapless phase and $y_0 = \infty$ in a massive phase.
For the case of $\mathbb{Z}_2$-symmetry the exponents are given by ($d=1$)
\begin{equation}
z=1 \, , \qquad \nu=1 \, , \qquad y_0 = \infty \, .
\label{expoZ2}
\end{equation}
With these identifications, our results (\ref{aboveHc}) and (\ref{belowHc})
for the isotropic $XY$ chain are consistent with the predictions from
general scaling arguments \cite{rosch}. Furthermore, eq.\ (\ref{dTdHcU1})
obeys the general scaling form
$(\partial T/\partial H)^{H=H_c}_S \propto T^{1 - 1/(\nu z)}$.

The situation with only $\mathbb{Z}_2$-symmetry is an exception in so far
as the general scaling arguments \cite{rosch} do not directly lead to
eq.\ (\ref{dTdHcZ2}). In this case, the leading term cancels at $H=H_c$ and
eq.\ (\ref{dTdHcZ2}) is obtained from the next order in $T$. For
$\mathbb{Z}_2$-symmetry, $H \ne H_c$ and $T \ll |H-H_c|$, the general
scaling predictions \cite{rosch} amount to
\begin{equation}
\frac{1}{T} \left(\frac{\partial T}{\partial H}\right)_S
  =\frac{1}{H-H_c} \, ,
\label{dTdHneqHc}
\end{equation}
which coincides with the result for the $U(1)$-symmetric system
in the gapped state eq.\ (\ref{aboveHc}) and agrees 
with an explicit computation for the
anisotropic $XY$ chain.

\section{Sawtooth chain}

\label{secSaw}

\begin{figure}[t]
\begin{center}
\includegraphics[width=0.5\columnwidth]{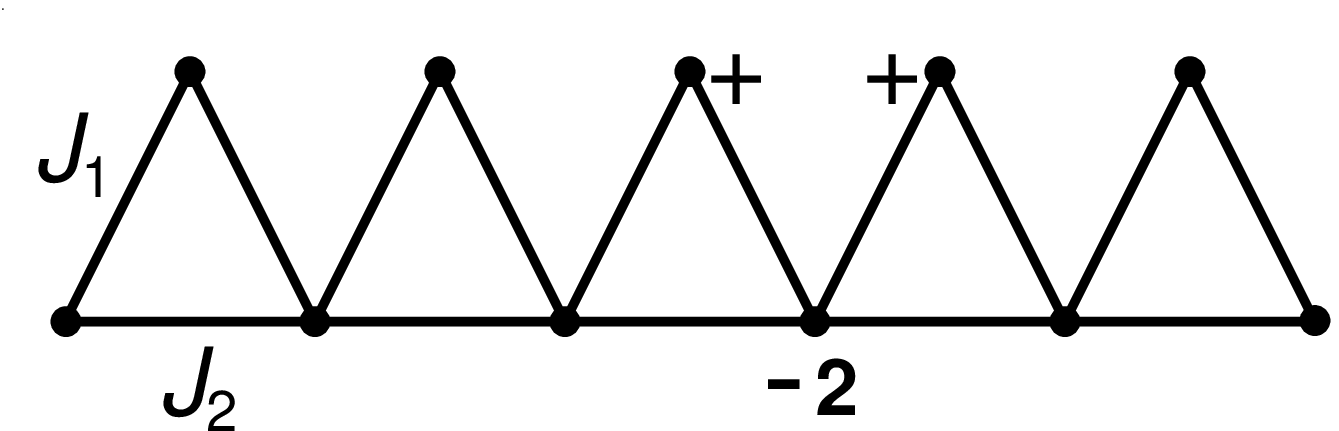}
\end{center}
\caption{\label{Sawtooth} 
The sawtooth chain. Localized magnons, which exist for $J_1=2J_2$,
are trapped within one valley between two adjacent triangles
with amplitudes indicated near each site.}
\end{figure}

The class of models with localized magnon excitations at the saturation field
includes the spin-1/2 Heisenberg models on kagome and pyrochlore lattices
\cite{schulenburg,richter,RSH}. In this section we study the
magnetothermodynamics of a one-dimensional representative of this class, 
namely the sawtooth chain  (called also a $\Delta$ chain).
The sawtooth chain is a chain of corner-sharing
triangles with two different exchanges for the base-base
and base-vertex edges, see Fig.~\ref{Sawtooth}. For the special 
ratio of the coupling constants $J_2=\frac{1}{2}J_1$ the sawtooth chain
supports localized magnon excitations, giving rise to a plateau in
the $T=0$ magnetization curve at half the saturation field followed by a jump
to saturation in the spin-1/2 model \cite{schulenburg,richter}.
Below we always assume the choice of parameters $J_2=\frac{1}{2}J_1$ with
$J_1\equiv J$.

In the fully saturated phase at high magnetic fields
the sawtooth chain has two branches of single spin-flip excitations
with energies
\begin{equation}
\varepsilon_{1k}=H-2J \ , \ \ \varepsilon_{2k}=H - \frac{J}{2}(1-\cos k) \ .
\end{equation}
Excitations from the low-energy dispersionless branch correspond
in real space to localized spin-flips residing in valleys between
two triangles, see Fig.~\ref{Sawtooth}. Such a localization
is a direct consequence of the frustration imposed by triangular
topology and resembles to a large extent similar behaviour
of magnons in kagome, pyrochlore and other frustrated lattices.
Localized magnons in the sawtooth chain condense below 
the saturation field $H_c=2J$ and form at zero temperature
a dense magnon crystal with one spin-flip occupying
every second valley between triangles \cite{schulenburg,richter}.
The magnon crystal ground state
is two-fold degenerate and breaks the translational symmetry.
In the following we discuss the low-temperature thermodynamics 
of the sawtooth
chain in the vicinity of the saturation field, which is dominated
by excitations from the lower branch. 

We consider a chain which consists of $L$ triangles ($N=2L$ spins). 
The lowest energy of a state with $n$ excited magnons, $n\leq L/2$, 
is equal to $n\varepsilon$, $\varepsilon=H-H_c\equiv\varepsilon_{1k}$.
The corresponding states are formed by nonoverlapping localized 
spin-flips, which  occupy spatially separated valleys.
This suggests a {\harddimer} representation of the lowest energy $n$-magnon
states so that the presence of a dimer on a given lattice
site of the basal chain in Fig.~\ref{Sawtooth} forbids
occupation of two neighbouring sites by other dimers.
Such a simple construction of the lowest energy states in every
$n$-magnon sector is complicated by the fact that the corresponding
states are not orthogonal due to an overlap
of the wave-functions of spin-flips trapped in two adjacent valleys.
The states are, however, linearly independent and can be properly
orthogonalized. Besides such lowest-energy localized states 
there are also scattering states in every $n$-magnon sector.
The scattering states of $n$ magnons are separated from 
$n\varepsilon$ by a finite gap  $\Delta_n$. For the two-magnon
sector of the sawtooth chain the variational calculations estimate 
$\Delta_2\approx 0.44J$ and suggest $\Delta_n=O(\Delta_2)$ in other
sectors \cite{mzh04}. The present numerical calculations confirm this picture 
yielding $\Delta_2\simeq 0.42J$, whereas for the magnon crystal state
on the 1/2-plateau $\Delta_{L/2}\simeq 0.25J$.
Neglecting at low temperatures exponentially small corrections
from the scattering states, we are left with the calculation of the free energy
of {\harddimers}. This is a classical problem
without dynamics and can be calculated exactly (see \cite{mzh04,baxter}).
The classical partition function of the {\harddimers} is
\begin{equation}
Z = \sum_{\{\sigma\}} 
\exp\left[-\frac{\varepsilon}{T}\sum_i\sigma_i\right]
\prod_{\langle ij\rangle} (1-\sigma_i\sigma_j) \ ,
\label{Zsaw}
\end{equation}
where the index $i$ runs over the sites of the basal chain and 
$\sigma_i=0,1$ denotes the absence or presence of a dimer on
a given site. The exclusion principle for nearest-neighbour
sites is imposed by the last term.
The partition function (\ref{Zsaw}) can be rewritten as
$Z = {\rm tr} {\cal T}^L$,  ${\cal T}$ being the individual transfer
matrix given by
\begin{equation}
{\cal T}(\sigma,\sigma') = \left(\begin{array}{cc}
1 & {\rm e}^{-\varepsilon/2T} \\
{\rm e}^{-\varepsilon/2T} & 0 \end{array}\right).
\end{equation}
This immediately yields the free energy and the entropy
normalized per triangle
\begin{eqnarray}
F/L & = & -T\ln\left(\frac{1}{2} + \sqrt{\frac{1}{4} + {\rm e}^{-\varepsilon/T}}
\right), \nonumber \\
S/L & = & \ln\left(\frac{1}{2}\!+\!
\sqrt{\frac{1}{4}+ {\rm e}^{-\varepsilon/T}}\right)
 + \frac{\varepsilon}{4T}\left( 
2 - \frac{1}{\sqrt{\frac{1}{4}+{\rm e}^{-\varepsilon/T}}} \right).
\label{HDsawEnt}
\end{eqnarray}
At $H=H_c$ the total entropy is 
\begin{equation}
S/L = \ln\left(\frac{1+\sqrt{5}}{2}\right) \approx 0.481212 \ ,
\label{ST0}
\end{equation}
which amounts to $34.7$\% of the total entropy $2\ln 2$ of the spin-1/2 
sawtooth chain. The same zero-temperature entropy of the sawtooth chain
at $H=H_c$  has been  independently obtained in \cite{DeRi}. It is also known to express
the entropy of an Ising antiferromagnetic chain at the saturation field
\cite{metcalf}. The entropy $S_{H=H_c}$ has no temperature dependence 
in the {\harddimer} approximation. 
A temperature dependence does arise from neglected
multimagnon states and its magnitude 
can test the validity of the {\harddimer} representation
for the low temperature thermodynamics of the sawtooth chain.

\begin{figure}[t]
\begin{center}
\includegraphics[width=0.7\columnwidth]{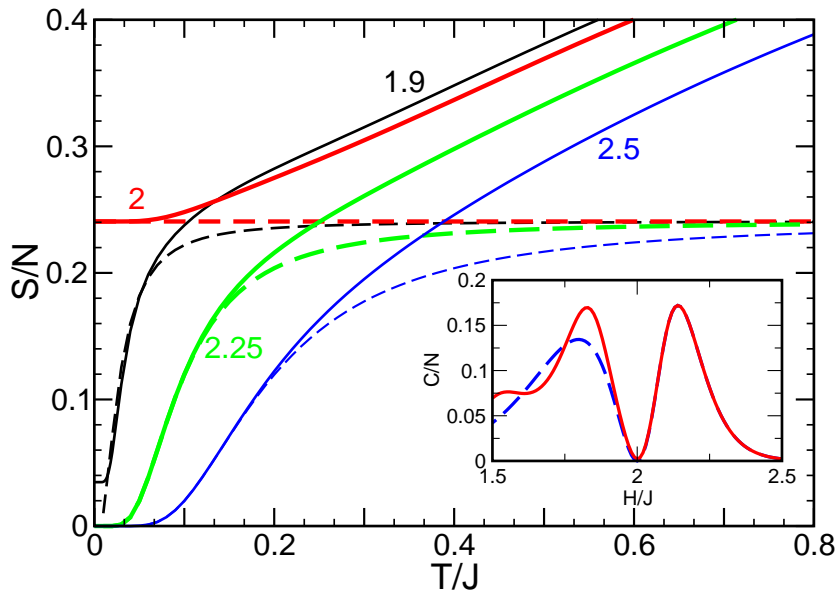}
\end{center}
\caption{\label{Sawthermo} 
The temperature dependence of entropy for the sawtooth chain
for different magnetic fields.
Full lines are numerical results for $N=20$ sites, 
dashed lines are from the {\harddimer} approximation. 
The number near each line indicates the corresponding value of the external
field. The inset shows the field dependence of the specific heat $C$ 
at $T=0.05J$ from numerical (full line) and  the {\harddimer}
(dashed line) calculations.}
\end{figure}

The results on the temperature dependence of
the entropy obtained by exact diagonalization for $N=20$ sites
and by the {\harddimer} mapping 
are compared in Fig.~\ref{Sawthermo} for several magnetic fields.  
For $H \ge H_c=2$, the numerical
results are in excellent agreement with the {\harddimer} approximation
at low temperatures consistent with a gap $\Delta \sim 0.2J$ to states
neglected in the {\harddimer} picture. The magnon crystal arising for
$H < H_c$ is two-fold degenerate. Hence, one obtains a non-extensive
entropy $S=\ln 2$ at $T=0$, giving rise to finite-size effects in $S/N$
for $H < H_c$ (see numerical curve for $H=1.9$ in Fig.~\ref{Sawthermo}). In this
regime, the {\harddimer} approach is superior at low temperatures since
it does not suffer from finite-size effects.
The inset of Fig.~\ref{Sawthermo} shows the field dependence of the specific
heat at $T=0.05J$. For $H \ge H_c$, the {\harddimer} approximation agrees
very well with the numerical results for $N=20$. For $H < H_c$, both approaches
become approximate since the numerical result (full line) is affected by
finite-size effects and the {\harddimer} approximation neglects states
that are relevant in this regime. The two peaks structure in the specific heat
on both sides of the critical field obtained in the two approaches 
is typical for field-transitions 
with finite entropy at $H=H_c$ \cite{mzh04}.

\section{Comparison of different degrees of frustration}

\label{secComp}

In the preceding sections we have considered the spin-1/2 Heisenberg model, 
the $J_1$--$J_2$ chain at $J_2 = \frac{1}{4}J_1$, and the sawtooth chain 
at $J_2 = \frac{1}{2}J_1$. These models can be considered
as examples for increasing degree of frustration, as is reflected e.g.\
by an increase of the entropy at low temperatures at the saturation
field $H_c$. In this section we will compare their entropy and
adiabatic (de)magnetization curves in more detail.

\begin{figure}
\begin{center}
\includegraphics[angle=270,width=0.7\columnwidth]{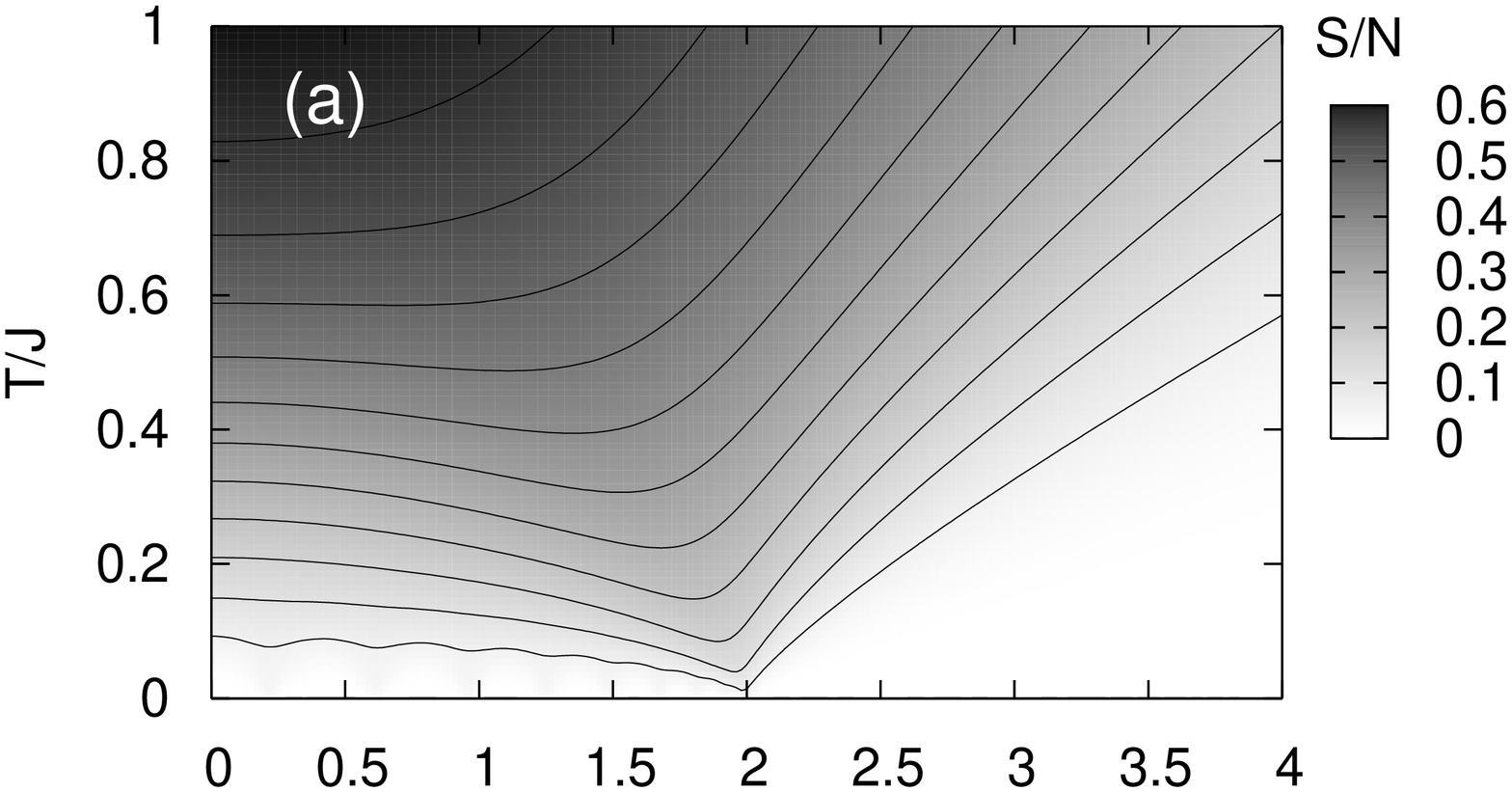}
\includegraphics[angle=270,width=0.7\columnwidth]{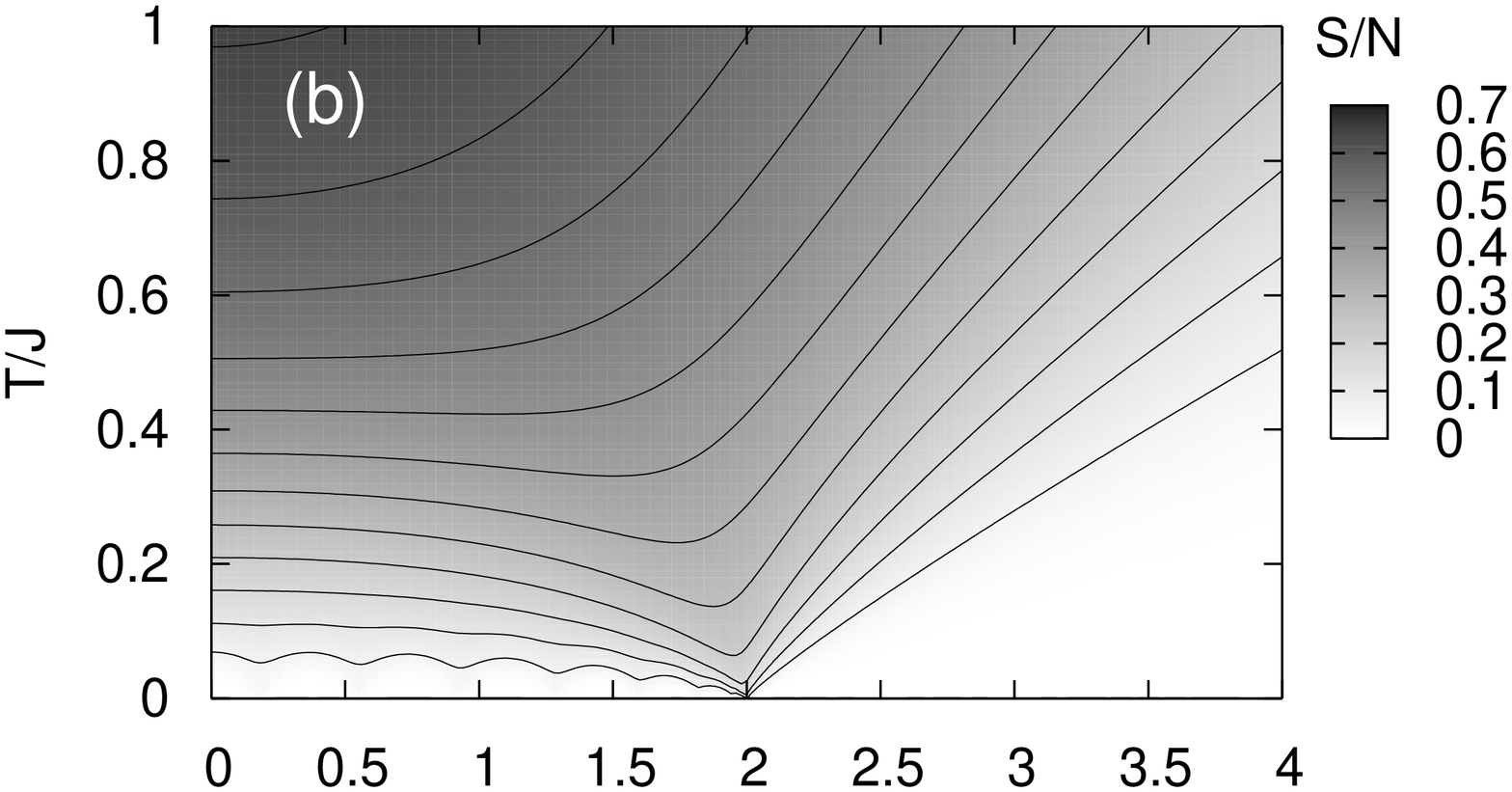}
\includegraphics[angle=270,width=0.7\columnwidth]{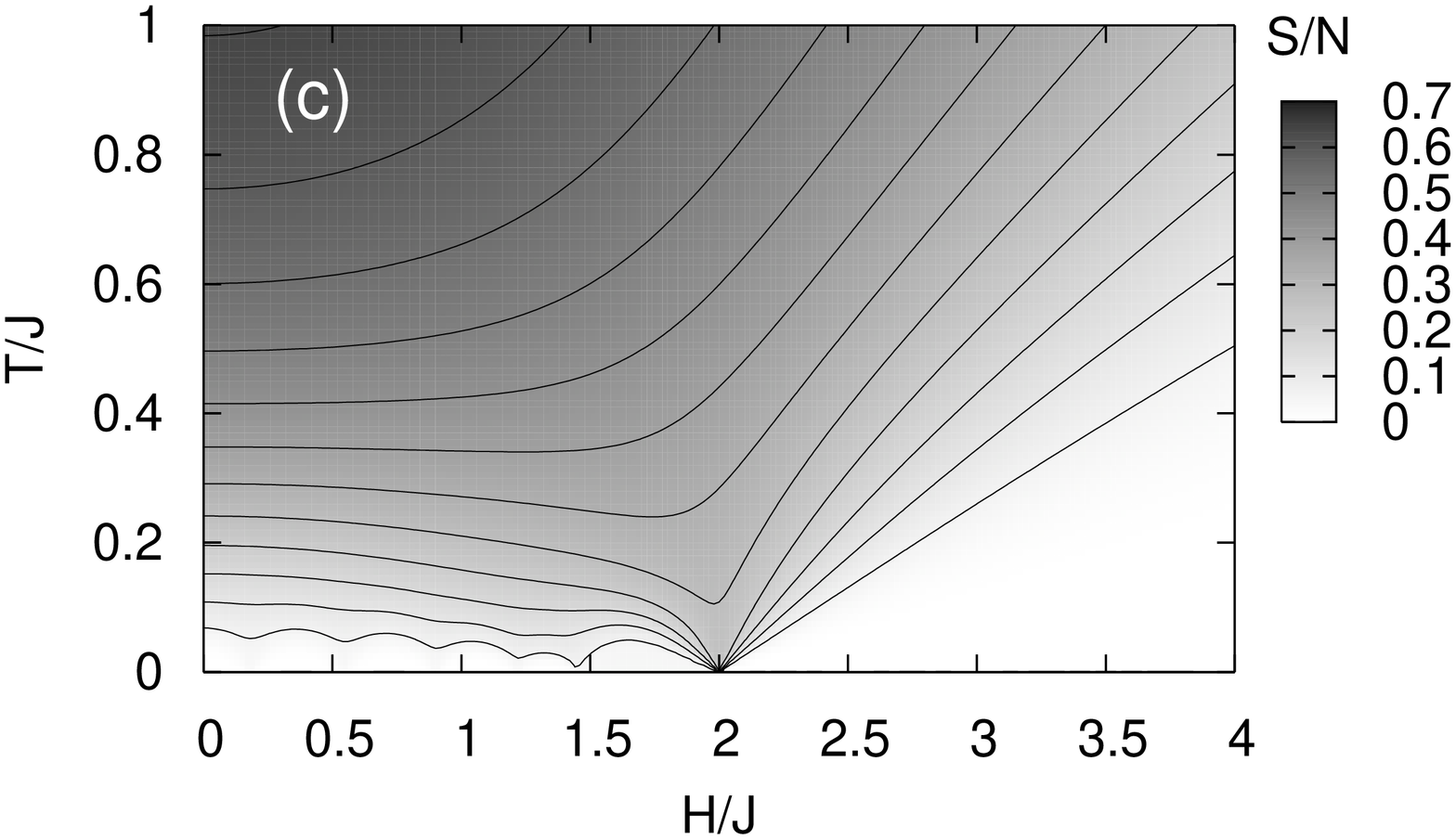}
\end{center}
\caption{\label{numEnt}
Entropy of $N=20$-sites lattices as a function of temperature and magnetic
field for spin-1/2 models:
(a) the Heisenberg chain,
(b) the $J_1$--$J_2$ chain with $J_1=J$, $J_2=\frac{1}{4}J$, and
(c) the sawtooth chain with $J_1=J$, $J_2=\frac{1}{2}J$.
Lines show curves of constant
entropy starting at the bottom with $S/N = 0.05$ and increasing in
steps of $0.05$.
}
\end{figure}

Fig.~\ref{numEnt} shows the entropy as a function of temperature and magnetic
field for the Heisenberg chain, the $J_1$--$J_2$ chain with $J_1=J$, 
$J_2=\frac{1}{4}J$, and the sawtooth chain with $J_1=J$, $J_2=\frac{1}{2}J$. 
The saturation field is $H_c=2J$ in all three cases. The contour lines show 
curves of constant entropy, {\it i.e.}\ adiabatic (de)magnetization curves.
The global behaviour of all three models exhibits some similarities,
in particular at high temperatures or sufficiently high above the
saturation field. The main difference occurs at low temperatures and
close to the saturation field, where the entropy clearly increases
from the Heisenberg chain (a) to the $J_1$--$J_2$ chain (b) and further
to the sawtooth chain (c) -- compare also Figs.~\ref{numHsat} and
\ref{Sawthermo} for more detailed data exactly at $H_c$.
The increase of entropy enhances the magnetocaloric effect as
the magnetic field is swept through saturation.

\begin{figure}[t]
\begin{center}
\includegraphics[width=0.7\columnwidth]{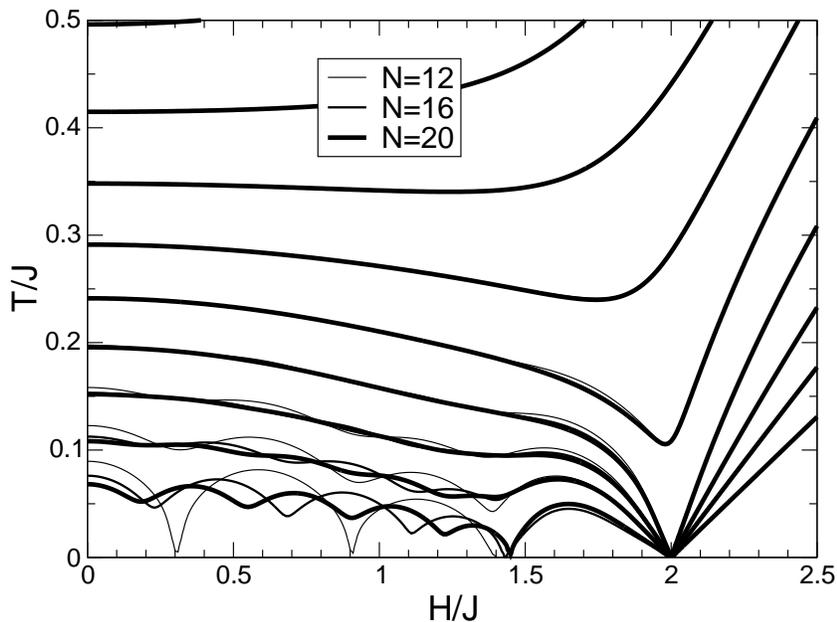}
\end{center}
\caption{\label{sawEnt}
Lines of constant entropy, {\it i.e.}\ adiabatic (de)magnetization
curves for the sawtooth chain with $J_1=J$, $J_2=\frac{1}{2}J$
for $N=12$, $16$ and $20$.}
\end{figure}

The data in Fig.~\ref{numEnt} has been computed numerically
for $N=20$ sites. Hence, it is important to discuss finite-size
effects. Some related comments for the entropy at the saturation field
have already been made in the context of Figs.~\ref{numHsat} and
\ref{Sawthermo}. We now examine finite-size effects for all fields,
focusing on the example of the sawtooth chain.
Fig.~\ref{sawEnt} corresponds to a part of
Fig.~\ref{numEnt}(c) at low temperatures and fields and shows
curves of constant entropy for $N=12$, $16$ and $20$. The
lowest curves start at $S/N = 0.05$ and increase in steps of $0.05$.
At low temperatures and close to $H_c = 2 J$ we can compare
with eq.\ (\ref{HDsawEnt}). The corresponding lines of constant entropy
are straight lines since the entropy of {\harddimers}
depends on $H$ and $T$ only via the combination $(H-H_c)/T$.
The lines obtained from eq.\ (\ref{HDsawEnt}) lie on top of the lines
obtained numerically for $N=20$ in Fig.~\ref{sawEnt} in the regime where
the latter are linear. Since eq.\ (\ref{HDsawEnt}) is valid for the
thermodynamic limit, this means that finite-size effects in Fig.~\ref{sawEnt}
are within the width of the lines for $N=20$ for $H$ close to $H_c$.
Further comparison of the curves for different system sizes shows that the
adiabatic demagnetization curves of a sawtooth chain with $N=20$ sites
are essentially free of
finite-size effects, except for $H \lesssim 1.7 J$, $T \lesssim 0.15J$.
These finite-size effects are related to the recovery of a gapless
spectrum at $T=0$ and for $H \lesssim 1.5 J$.
Similarly, finite-size effects can also be neglected for the Heisenberg
chain and the $J_1$--$J_2$ chain (Fig.\ \ref{numEnt}(a) and (b))
except for $H<2 J$, $T \lesssim 0.15J$.
Note that in the gapless regimes ($H < H_c$) temperature varies only weakly
with the applied field (compare also Fig.\ \ref{XYad}).

\begin{figure}[t]
\begin{center}
\includegraphics[width=0.7\columnwidth]{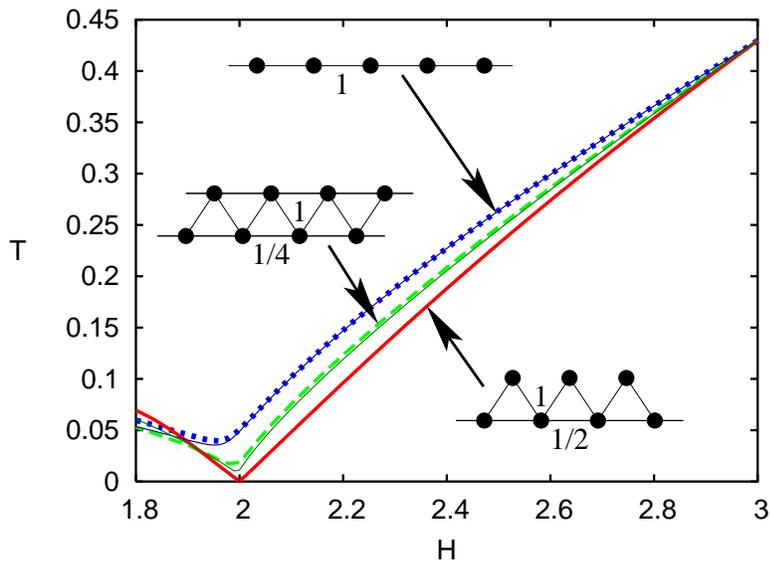}
\end{center}
\caption{\label{mcompFig}
Comparison of adiabatic demagnetization curves when approaching
the saturation field from above for the spin-1/2 Heisenberg chain,
the $J_1$--$J_2$ chain with $J_1 = 1$, $J_2 = 1/4$, and
for the sawtooth chain with $J_1=1$, $J_2=1/2$, as indicated
in the figure. The three bold curves have been computed numerically
for $N=20$ sites, the two thin curves have been obtained in a 
free-fermion approximation.}

\end{figure}

Finally, Fig.~\ref{mcompFig} presents a direct comparison of
adiabatic demagnetization curves for the spin-1/2 Heisenberg chain,
the $J_1$--$J_2$ chain and the sawtooth chain. We start
with all systems at the same temperature and a magnetic field
noticeably above the saturation field ($H=3$). During adiabatic
demagnetization, all three systems cool upon approaching the
saturation field. Clearly, the $J_1$--$J_2$ chain cools down to lower
temperatures than the spin-1/2 Heisenberg chain under the same
conditions, and the sawtooth chain to even lower temperatures.
Along with numerical results we show by thin lines the analytic results
for a free-fermion model for the Heisenberg chain and the $J_1$--$J_2$
chain. One observes that free fermions yield an excellent description
of the Heisenberg chain above the saturation field while slightly
bigger deviations for the $J_1$--$J_2$ chain are due to the relevance
of interactions in the latter case.

The adiabatic demagnetization curve in Fig.~\ref{mcompFig}
of the sawtooth chain corresponds to an entropy below the value given
in eq.~(\ref{ST0}). Hence, one even observes cooling to $T \to 0$
as $H \to H_c = 2$.
Nevertheless, the example of Fig.~\ref{mcompFig} is still
at comparably high temperatures (entropy). Among others, this ensures
that finite-size effects can be neglected although the bold curves in
Fig.~\ref{mcompFig} have been computed with $N=20$. While asymptotic
low-temperature expressions like eq.~(\ref{eqTf}) are not accurate
in this regime, the enhancement of the magnetocaloric effect 
with increasing frustration is evident.

\section{Conclusions and outlook}

When discussing the magnetocaloric effect at low temperatures, one should 
clearly distinguish two different cases: (i) large relative 
variations of entropy in a magnetic field, while the total entropy tends to
zero, and (ii) large absolute variations of entropy. Let us first 
discuss the former case, which appears in ordinary nondegenerate quantum spin 
chains. We have shown that an enhanced magnetocaloric effect
exists in the vicinity of quantum phase transitions in a magnetic field.
Specifically, we have studied the transition at the saturation field.
There are, however, many other systems/transitions to which our predictions
apply, in particular second-order transitions at the boundaries of
magnetization plateaux \cite{RSH,MiKo}. Field-induced transitions in 1D
systems with a singlet ground state, like spin ladders or Haldane gap chains,
are one specific example (see e.g.\ \cite{MiKo} for a recent review).
Triplet excitations with $\Delta S^z=+1$ condense, when the Zeeman energy 
exceeds the triplet gap. Triplets can be
represented as hard-core bosons, which are mapped in one dimension to weakly
interacting spinless fermions \cite{takahashi,sachdev}.
Therefore, the entropy of triplet excitations close to the transition
field is also described by the
equations (\ref{eqSxy}) in the single-particle approximation.
A less trivial example is the Bose condensation of magnetic excitations 
in 2D spin systems with 1D type degeneracies in the energy spectra.
Such a possibility occurs in the $J_1$--$J_2$ antiferromagnetic model
on a square lattice near the saturation field \cite{jackeli}
as well as in the 2D singlet gap magnet Cs$_3$Cr$_2$Br$_9$ \cite{cscrbr}.
Triplet excitations above the singlet gap 
have also a large degeneracy in the singlet 
state of SrCu$_2$(BO$_3$)$_2$ \cite{kageyama,totsuka}
which should lead to a large entropy release at the 
gap closing transition. 
In the frustrated antiferromagnetic chain
the magnetocaloric effect is enhanced  due to 
a cancellation of the leading $q^2$ term in the one-particle dispersion (\ref{frus1P}) 
for $J_2/J_1=4$. Such a cancellation remains valid for ferromagnetic
sign of the nearest-neighbour exchange $J_1<0$.
The magnetization curve of the $J_1$--$J_2$ chain with ferromagnetic $J_1<0$
is also particularly steep below the saturation 
(see e.g.\ Fig.\ 4 of \cite{zigzag}), suggesting that the magnetocaloric effect
may also be enhanced in this regime. Rb$_2$Cu$_2$Mo$_3$O$_{12}$ is
a candidate for a realization in the latter parameter regime with a moderate
saturation field \cite{hase}.

A preliminary experimental indication of a large magnetocaloric effect 
has been obtained in pulsed field measurements in a spin-1/2 Heisenberg  
chain antiferromagnet \cite{wolf}. In the vicinity of the saturation field 
the magnetization curve deviates significantly from the behaviour 
expected for the corresponding bath temperature indicating a sizeable 
magnetocaloric effect. Thus, the magnetocaloric effect can significantly
affect experimental results obtained in pulsed magnetic fields.
A systematic experimental investigation of the magnetocaloric
effect near quantum phase transitions can also be used with the help of
eqs.~(\ref{dTdHcZ2}), (\ref{dTdHcU1}) to clarify the role of spin
anisotropies in a given magnetic material.

Large absolute variations of entropy in a magnetic field
appear for a class of geometrically frustrated antiferromagnetic
models. The sawtooth chain considered in sections \ref{secSaw} and
\ref{secComp} has a finite zero-temperature entropy at the saturation field.
The analysis of the magnetothermodynamics of this model shows
that in a strongly correlated spin system there are
emergent ``paramagnetic'' degrees of freedom, which experience an effective
magnetic field $h=(H-H_c)$. Such effective ``paramagnetic moments''
are responsible for an enhanced cooling
power in the vicinity of $H_c$ very similar to standard
paramagnetic salts \cite{tishin}.
However, the ``paramagnetic'' language is not completely adequate
and a picture of hard localized magnons is more appropriate for
the physical description.
Among the differences with a simple paramagnetic picture, we mention
the geometrical origin of the finite entropy: the zero-$T$
entropy at the saturation field
does not depend on the value of the local spin. Also, there is an asymmetry
between the $h>0$ and $h<0$ regions close to the saturation field
in Fig.~\ref{sawEnt}.

The antiferromagnetic sawtooth chain model is 
realized in delafossite YCuO$_{2.5}$ \cite{delafossite},
though with a yet unknown ratio of the exchange constants $J_2/J_1$.
Our theoretical study should also be relevant for
2D and 3D spin models on kagome, pyrochlore and garnet (hyper-kagome)
lattices, which have numerous experimental realizations.
The magnetocaloric effect has been known for a long time
in gadolinium gallium garnet Gd$_3$Ga$_5$O$_{12}$, which is often 
utilized in adiabatic demagnetization refrigerators
\cite{brodale,barclay}.
Very recently, a large magnetocaloric effect has also been measured in
the pyrochlore antiferromagnet Gd$_2$Ti$_2$O$_7$ \cite{sosin}.
Magnetic Gd$^{3+}$ ions have a rather large spin $S=7/2$. This
justifies the classical approximation applied for 
pyrochlore and garnet antiferromagnets in \cite{mzh03}.
Still, the classical model cannot describe the
behaviour at low temperatures $T\lesssim JS$.
One of the physical implications of the present work 
is that the entropy on frustrated lattices has a purely geometrical
origin. Therefore, $S=1/2$ frustrated magnetic materials should be as good
as $S\gg 1/2$ compounds for cooling by adiabatic demagnetization,
contrary to the experience with paramagnetic salts \cite{tishin}.
We suggest that a synthesis of new spin-1/2 
antiferromagnetic materials with kagome, pyrochlore and garnet lattices 
should be interesting both  from a fundamental point of view
and with regard to possible technological applications
in the field of low-temperature magnetic cooling.

\ack
We would like to thank G. Jackeli, A.\ Kl\"umper, J.\ Richter,
A.\ Rosch, H.\ Tsunetsugu, and S.\ Wessel  for useful discussions.
The more complicated numerical computations presented in this article have
been performed on the compute-server {\tt cfgauss} at the computing center
of the TU Braunschweig. A.H.~would
like to acknowledge the hospitality of the Institute for Theoretical Physics
of the University of Hannover during the course of this work.

\end{document}